\documentclass[fleqn,twoside]{article}
\usepackage{espcrc2}

\usepackage{graphicx}

\newcommand{\AmS}{{\protect\the\textfont2
  A\kern-.1667em\lower.5ex\hbox{M}\kern-.125emS}}
\newcommand{\bp}{{\bf b}_\perp}
\newcommand{\yT}{{\bf y}_\perp}
\newcommand{\xT}{{\bf x}_\perp}
\newcommand{\oT}{{\bf 0}_\perp}

\newcommand{\kT}{{\bf k}_\perp}

\newcommand{\bi}{\begin{itemize}}
\newcommand{\ei}{\end{itemize}}
\newcommand{\be}{\begin{eqnarray}}
\newcommand{\ee}{\end{eqnarray}}
\newcommand{\bea}{\begin{eqnarray}}
\newcommand{\eea}{\end{eqnarray}}

\title{Quark correlations and single-spin asymmetries}

\author{M. Burkardt\address[MCSD]{Department of Physics,
New Mexico State University,\\ 
Las Cruces, NM 88003, U.S.A.}%
        \thanks{This work was supported by the Department of Energy 
(DE-FG03-96ER40965)
} }
       
\begin{document}

\begin{abstract}
We illustrate, in a semi-classical picture, how the Wilson line 
phase factor in gauge invariantly defined unintegrated parton density
can lead to a nonzero single-spin asymmetry (Sivers effect).
\vspace{1pc}
\end{abstract}

\maketitle

\section{INTRODUCTION}

Many hadron production reactions give rise to significant
$\perp$ single-spin asymmetries (SSA) which persist at high energies.
For example, hyperons produced
in hadronic collisions typically show a large polarization
perpendicular to the production plane.
More recently, the HERMES collaboration found a significant
SSA 
(left-right asymmetry in the transverse momentum distribution
of the produced mesons
in the directions perpendicular to the nucleon spin)
in semi-inclusive production of $\pi$ and $K$ mesons \cite{hermes}.
Theoretically, two mechanisms have been proposed to explain
the asymmetry: the Sivers mechanism, where the final state 
interactions (FSI) give rise to a $\perp$ momentum asymmetry
of the active quark already before it fragments into hadrons 
\cite{sivers}, and
the Collins mechanism, where the asymmetry arises when a $\perp$
polarized quark fragments into mesons (see Ref. \cite{collins2}
and references therein).
Both mechanisms have recently been observed
by the HERMES collaboration (they can be disentangled by also
measuring the scattering plane of the electron) \cite{hermes}.

Simple model calculations \cite{BHS,diquarkmodel} 
have revealed that, even at high energies, the
FSI can indeed give rise to a non-vanishing transverse 
single-spin asymmetry for the active quark 
(Sivers mechanism). Those FSI can formally be incorporated 
into a 
definition of unintegrated parton densities by introducing
appropriate Wilson line phase factors \cite{ji,collins}.

The Sivers mechanism is interesting for a variety of reasons
\bi
\item it vanishes under naive time-reversal and a non-trivial
phase from the FSI is needed to give a nonzero effect
\item[$\hookrightarrow$] the Sivers mechanism provides 
information about the space-time structure of the target
\item A nonzero Sivers mechanism implies a nonzero
Compton amplitude involving nucleon helicity flip
without quark helicity flip
\item[$\hookrightarrow$] requires (nonperturbative) helicity
nonconservation in the nucleon state ($\chi$SB!)
\item in model calculation the mechanism requires interference
between phases of wave function components that differ by
one unit of orbital angular momentum
\item[$\hookrightarrow$] the effect may provide novel
insights about orbital angular momentum in the proton.
\ei
The paper is organized as follows: in Section 2 we will
introduce gauge invariant unintegrated parton densities.
in Section 3, we will discuss the physics associated with the
Wilson line gauge links in gauge invariantly defined 
parton densities, and in Section 4 we will investigate the
asymmetry in light-cone
gauge and explain how the Sivers mechanism can be related to
quark correlations in the $\perp$ plane.
We will then conclude with a simple semi-classical picture
that relates $\perp$ deformations of the quark densities with
the average sign of the Sivers mechanism.

\section{UNINTEGRATED PARTON DENSITIES}
The naive definition of a
unintegrated (i.e. $\kT$-dependent) parton density reads
\be
q_{naive}(x,{\bf k}_\perp)
\!\!\!&=& \!\!\!\!\!\int \frac{dy^- d^2\yT }{16\pi^3}e^{-ixp^+y^-+i\kT \cdot \yT } \nonumber
\\&\times &\!\!\!\!
\left\langle p,s \left|
\bar{q}(0) \gamma^+ q(y)
\right| p,s\right\rangle. \label{naive}
\ee
This density is not only gauge dependent but also leads to
a vanishing asymmetry
$\int d^2\kT q_{naive}(x,{\bf k}_\perp)\kT=0$.
due to time-reversal invariance.

At first one might be tempted to render Eq. (\ref{naive})
gauge invariant by connecting the quark field operators
by a straight line gauge string (Wilson line), 
but this does not change
the time-reversal argument and the resulting asymmetry is
still zero. More importantly, the choice of path for the
gauge string is not arbitrary but should be determined by the
physical quantity that is observed in the experiment. 
If one is interested in the transverse momentum
of the knocked out quarks (i.e. mesons after fragmentation)
then the choice of path should be such that it reflects
the FSI of the outgoing quark. Since the active quark in DIS
is ultra-relativistic, the correct way to render Eq. (\ref{naive})
gauge invariant is by including Wilson lines from the position
of the active quark to infinity along the light-cone.
In addition, the two Wilson lines need to be connected at light-cone
infinity. The choice of path for the segment at $x^-=\infty$
is arbitrary as long as the gauge field  at $x^-=\infty$
is pure gauge. These considerations suggests the following
definition for unintegrated parton densities relevant for
semi-inclusive DIS
\be
q(x,\kT)\!\!\!\!&=& 
\!\!\!\!\!\!\int \frac{dy^- d^2\yT }{16\pi^3}e^{-ixp^+y^-+i\kT \cdot \yT } \label{correct}
\\&\times &\!\!\!\!\!
\left\langle p,s \left| \bar{q}_U(y)
U_{\left[\infty^-,\yT,\infty^-,\oT\right]}\gamma^+q_U(0)
\right|p,s\right\rangle .
\nonumber
\ee
with
\be
q_U(0)&\equiv& U_{\left[\infty^-,\oT;0^-,\oT\right]} q(0)\\
\bar{q}_U(y)&\equiv&\bar{q}(y) 
U_{\left[y^-,\yT;\infty^-,\yT \right]} \nonumber .
\ee
The $U$'s are Wilson line gauge links, for example
\be
U_{[0;\xi]}= P\exp\left(ig\int_0^1 ds \xi_\mu A^\mu(s\xi)\right)
\ee
connecting the points $0$ and $\xi$.
\begin{figure}
\unitlength1.cm
\begin{picture}(10,3.5)(3.65,15)
\includegraphics{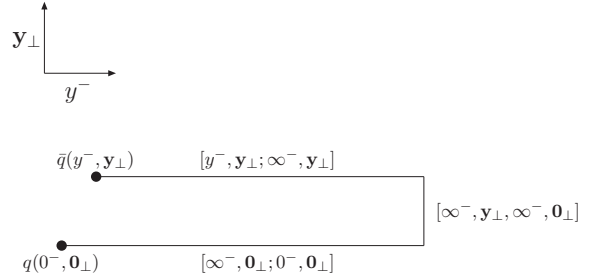}
\end{picture}
\caption{Illustration of the gauge links in gauge invariant
Sivers distributions (\ref{correct}).}
\label{fig:staple}
\end{figure}
With the gauge links to $x^-=\infty$ included, time 
reversal invariance no longer implies a vanishing asymmetry.
Indeed, under time reversal the direction of the gauge link 
changes and the only consequence of time reversal invariance
for Eq. (\ref{correct})
is opposite signs for the asymmetries in semi-inclusive pion 
production and Drell-Yan experiments \cite{collins2}.

While in principle all three gauge links in Fig. \ref{fig:staple}
contribute, the gauge link segment at $x^-=\infty$
is important only in light-cone gauge \cite{ji}. 

However, while the introduction of Wilson line phase
factors in gauge invariant parton densities has helped
to understand why there can  be a nonzero Sivers mechanism
at high energies, it has at the same time made the underlying
physics of the mechanism rather obscure: 
the asymmetry arises from interference between 
phase factors from different partial waves and in addition 
require a additional nontrivial phase
contribution from the Wilson line.
In the rest of these notes we will illustrate how
the Wilson line, together with the nucleon ground state
wave function, conspires to provide a transverse asymmetry.

\section{PHYSICS OF THE WILSON LINE PHASE FACTOR}

In order to illustrate the physics of the Wilson line factors,
we focus now on the average transverse momentum for quarks
of flavor $q$ 
\be
\langle {\bf k}_{\perp q}\rangle \equiv \int dx \int d^2\kT
q(x,\kT) \kT.
\ee
We evaluate $\langle {\bf k}_{\perp q}\rangle$ from Eq. 
(\ref{correct}) and integrate by parts, yielding 
\be
\langle {\bf k}_{\perp q}\rangle &\propto&\\
& &\!\!\!\!\!\!\!\!\!\!\!\!\!\!\!\!\!\!\!\!\!\!\!\!\!\!\!\!\!\!\!\!
\left.\frac{\partial}{\partial y_\perp}
\left\langle p,s \left|
\bar{q}_U(0) U_{\left[\infty^-,\yT,\infty^-,\oT\right]}
\gamma^+ q_U(y)
\right| p,s\right\rangle\right|_{y=0}\!\!\!\!\!\!\!\!\!\!\!\!\!\!\!\!
\nonumber
\ee
In this expression, the derivative can act either on the quark
field operator or the gauge links. The term where the derivative
acts on the quark field operator vanishes again due to time
reversal invariance [it corresponds to the asymmetry that one
would obtain starting from Eq. (\ref{naive})]. The interesting
term is obtained by acting on the gauge links, which yields
(modulo light-like gauge links) after some algebra 
\cite{qiu,mank,boer}
\be
\langle {\bf k}_\perp\rangle &\propto& \label{twist3}\\
& &\!\!\!\!\!\!\!\!\!\!\!\!
\left.\left\langle p,s \left|
\bar{q}(0) \gamma^+ \int_0^\infty d\eta^-  
G^{+\perp}(\eta)
q(0) \right| p,s\right\rangle\right.,
\nonumber
\ee
where terms that vanish because of time-reversal invariance have
been dropped. Here $G^{\mu \nu}$ is the gluon field strength tensor.

This result has a simple semi-classical interpretation:
$G^{+\perp}(\eta)$ is the $\perp$ component of the
force from the spectators on the active quark.
Integrating this force along the trajectory 
(for a ultrarelativistic particle, time=distance)
of the outgoing quark then yields the $\perp$ impulse
 $\int_0^\infty d\eta^-  G^{+\perp}(\eta)$ which the active
quarks acquires from the FSI as it escapes the hadron.
The average $\kT$ is then obtained by correlating the
quark density with the $\perp$ impulse.

However, although this result nicely illustrates the physics of the
contribution from the Wilson lines, it still does not tell us the
sign/magnitude of the asymmetry. Indeed, early estimates for 
Eq. (\ref{twist3}) concluded that the resulting asymmetry should
be very small \cite{mank}.

\section{SIVERS MECHANISM IN $A^+=0$ GAUGE}
There are several reasons why one is interested to proceed with
light-cone gauge $A^+=0$. First of all this is the most physical
gauge for a light-cone description of hadrons, which is in turn
the natural framework to describe DIS. Secondly, all light-like
Wilson lines become trivial in this gauge. Furthermore,
there exists already a rich phenomenology for light-cone
wave functions of hadrons.

However, if one neglects the gauge link at $x^-=\infty$ in Eq. 
(\ref{correct}) then the $\perp$ asymmetry vanishes in $A^+=0$ gauge:
as we mentioned above the light-like gauge links are trivial in
this gauge and without the gauge link at $x^-=\infty$, Eq. 
(\ref{correct}) reduces to Eq. (\ref{naive}) which yields a
vanishing asymmetry due to time-reversal invariance.
In fact, as was revealed by explicit model calculations in Ref.
\cite{BHS}, very careful regularization of the light-cone
zero-modes (see Ref. \cite{advances} and references therein) 
is required if one wants to calculate the asymmetry in $A^+=0$ 
gauge.  

Starting again from Eq. (\ref{correct}) and setting $A^+=0$
one finds
\be
\langle {\bf k}_{\perp q}\rangle &\propto&
\label{lc1}\\
& &\!\!\!\!\!\!\!\!\!\!\!\!\!\!\!\!\!\!\!\!\!\!\!\!\!\!\!\!\!\!\!\!
\left.\frac{\partial}{\partial y_\perp}
\left\langle p,s \left|
\bar{q}(0) U_{\left[\infty^-,\yT,\infty^-,\oT\right]}
\gamma^+ q(y)
\right| p,s\right\rangle\right|_{y=0}\!\!\!\!\!\!\!\!\!\!\!\!\!\!\!\!
\nonumber\\
& &\!\!\!\!\!\!\!\!\! =ig\left\langle p,s \left|
\bar{q}(0){\bf A_\perp}(\infty^-,\oT)q(0)
\right| p,s\right\rangle \nonumber .
\ee
\subsection{Finiteness conditions}
In order to make further progress, we need to express the gauge
field at $x^-=\infty$ in terms of less singular
degrees of freedom at finite $x^-$. We will do this in several
steps. First me make use of the time reversal invariance and
replace ${\bf A_\perp}(\infty^-,\oT)$ by $\frac{1}{2}\left[
{\bf A_\perp}(\infty^-,\oT) -{\bf A_\perp}(-\infty^-,\oT)\right]$
in Eq. (\ref{lc1}). Then we use the fact that the gauge field
at $x^-=\pm \infty$ must be pure gauge, i.e. we impose as a 
condition on the states
\be
G^{+-}(x^-\!\!=\pm\infty,\xT)= G^{12}(x^-\!\!=\pm\infty,\xT)=0
\label{gauge}
.\ee
In light-cone gauge, $G^{+-}_a=\partial_-A^-_a$, and therefore
Eq. (\ref{gauge}) implies $\partial_-A^-(x^-=\pm\infty,\xT)=0$. 
Integrating the constraint equation for $A^-$
\be
-\partial_-^2 A^-_a -\partial_-\partial^i A^i_a - gf_{abc} 
A^i_{b} G^{i+}_c = j^+_a,
\ee
we thus find the first finiteness condition \cite{qsivers}
\be
\partial^i \alpha^i_a(\xT) = -\rho_a(\xT),
\ee
where
\be
\alpha^i(\xT) = \frac{1}{2}\left[
A^i_a(\infty^-,\xT) - A^i_a(-\infty^-,\xT)\right]
\ee
is the anti-symmetric piece of the $\perp$ gauge field at
the boundary and
\be
\rho_a(\xT) = g\int dx^-\left[ \bar{q} \gamma^+ 
\frac{\lambda^a}{2}q  + f_{abc} A^i_{b} G^{i+}_c \right]
\ee
is the total (quark+glue) color charge density integrated
along $x^-$. If one imagines the Lorentz contracted proton
as a pizza then $\rho_a(\xT)$ is the color charge density
operator at position $\xT$ on the pizza.

Finally, imposing 
$G^{12}(x^-=\pm\infty,\xT)=0$ implies
\be
A^j(\infty^-,\xT)=\frac{i}{g}U^\dagger(\xT)\partial^j U(\xT)
\ee
and similarly at $x^-=-\infty$ with a different $U$.

It is instructive to solve these constraints perturbatively. To
lowest order one finds the ``abelian'' solution
\be
\alpha_a^i(\xT)
= -\int \frac{d^2\yT}{2\pi} \frac{x^i-y^i}{\left|\xT-\yT\right|^2}
\rho_a(\yT)
\ee
corresponding to a Lorentz boosted Coulomb field integrated along 
the $\hat{z}$-axis. Inserting this result into (\ref{lc1}) yields
\be
\label{corel}
\left\langle k^i_q \right\rangle &=& 
-\frac{g}{4p^+}
\int \frac{d^2\yT}{2\pi} \frac{y^i}{\left|\yT\right|^2}
\\
& &\times
\left\langle p,s \left|
\bar{q}(0)\gamma^+ \frac{\lambda_a}{2}q(0) \rho_a(\yT) 
\right| p,s \right\rangle ,\nonumber
\ee
which has again a very physical interpretation:
the average $\kT$ is obtained by summing over the $\perp$ impulse
caused by the color-Coulomb field (since we solved the constraint
equations only to lowest order) of the spectators.

One immediate consequence of this result is that the total
Sivers effect (for the gluon Sivers effect see Refs. 
\cite{rodrigues,vogel})
summed over all quarks and gluons with equal
weight is zero
\be
\sum_{c=q,g}\left\langle k^i_c \right\rangle=0
\label{zero}
\ee 
(by symmetry). One can show \cite{gsivers} that this result holds
beyond lowest order in perturbation theory. It should emphasized
that Eq. (\ref{zero}) is {\sl not} a trivial consequence of
momentum conservation since $\kT$ in the Sivers effect is {\sl not} 
the momentum of the partons before the collision
(which also enters the N\"other momentum).
Instead the $\perp$ momentum in the Sivers effect is the sum
of the momentum the partons have before being ejected {\sl plus}
the momentum they acquire due to the FSI. Since the momenta before
the partons are ejected add up to zero, Eq. (\ref{zero}) is thus
a statement about the net momentum due to the FSI: the net (summed
over all partons) $\perp$ momentum due to the FSI is zero, which is
a nontrivial result since what one adds up here is {\sl not}
the $\perp$ momenta of all fragments in the target but only the
$\perp$ momenta in the current fragmentation region.
Eq. (\ref{zero}) is therefore a nontrivial and useful constraint
on parameterizations of Sivers distributions
\cite{vogel,gsivers,ansel}.

Eq. (\ref{corel}) is also very useful for practical evaluation
of the Sivers effect from light-cone wave functions. The
original expression (\ref{lc1}) involved the gauge field at
$x^-=\infty$, which is very sensitive to the regularization
procedure, we have succeeded to express the asymmetry in 
terms of degrees of freedom at finite $x^-$, i.e. $\perp$
color density-density correlations in the $\perp$ plane.
Eq. (\ref{corel})
can be directly applied to light-cone wavefunctions, without further
regularization.

If we want to proceed further, we need a model for the light-cone
wave function. Here we do not want to consider a specific model,
but rather the whole class of valence quark models, which may be
useful for intermediate and larger values of $x$. In a valence quark
model, since the color part of the wave function factorizes,
one can replace the color density-density correlations by
neutral density-density correlations
\be
& &\\
\left\langle \bar{q}(0)\gamma^+ \frac{\lambda_a}{2}q(0) 
\rho_a(\yT) \right\rangle\!\!\!\!
&=& \!\!\!\!- \frac{2}{3} \left\langle \bar{q}(0)\gamma^+  q(0) 
\rho(\yT) \right\rangle \!\!\!\!\!\!\!\!\!\!\nonumber
\ee
and therefore
\be
\left\langle k^i_q \right\rangle &=& \frac{g}{6p^+}
\int \frac{d^2\yT}{2\pi} \frac{y^i}{\left|\yT\right|^2}
\label{corel2}\\
& & \times
\left\langle p,s \left|
\bar{q}(0)\gamma^+ q(0) \rho(\yT) 
\right| p,s \right\rangle\!\!\!\!\!\!\!\!\!\!
\nonumber
\ee
with $\rho(\yT)=\sum_{q^\prime} \int dy^-
\bar{q}^\prime(y^-,\yT)\gamma^+ q^\prime(y^-,\yT)$.

\subsection{Connection with GPDs}
From studies of generalized parton distributions (GPDs) it is
known that the distribution of partons in the $\perp$ plane 
$q(x,\bp)$ is
significantly deformed for a transversely polarized target 
\cite{IJMPA}.
The mean displacement of flavor $q$
($\perp$ flavor dipole moment) is given by
\be
\label{dipole}
d^q_y &\equiv& \int\!\! dx\!\int \!\!d^2\bp
q(x,\bp) b_y\nonumber\\ &=& 
\frac{1}{2M} 
\int\!\! dx E_q(x,0,0) = \frac{\kappa_{q}^p}{2M} .
\ee
\begin{figure}
\unitlength1.cm
\begin{picture}(10,9.5)(2.2,8.2)
\includegraphics{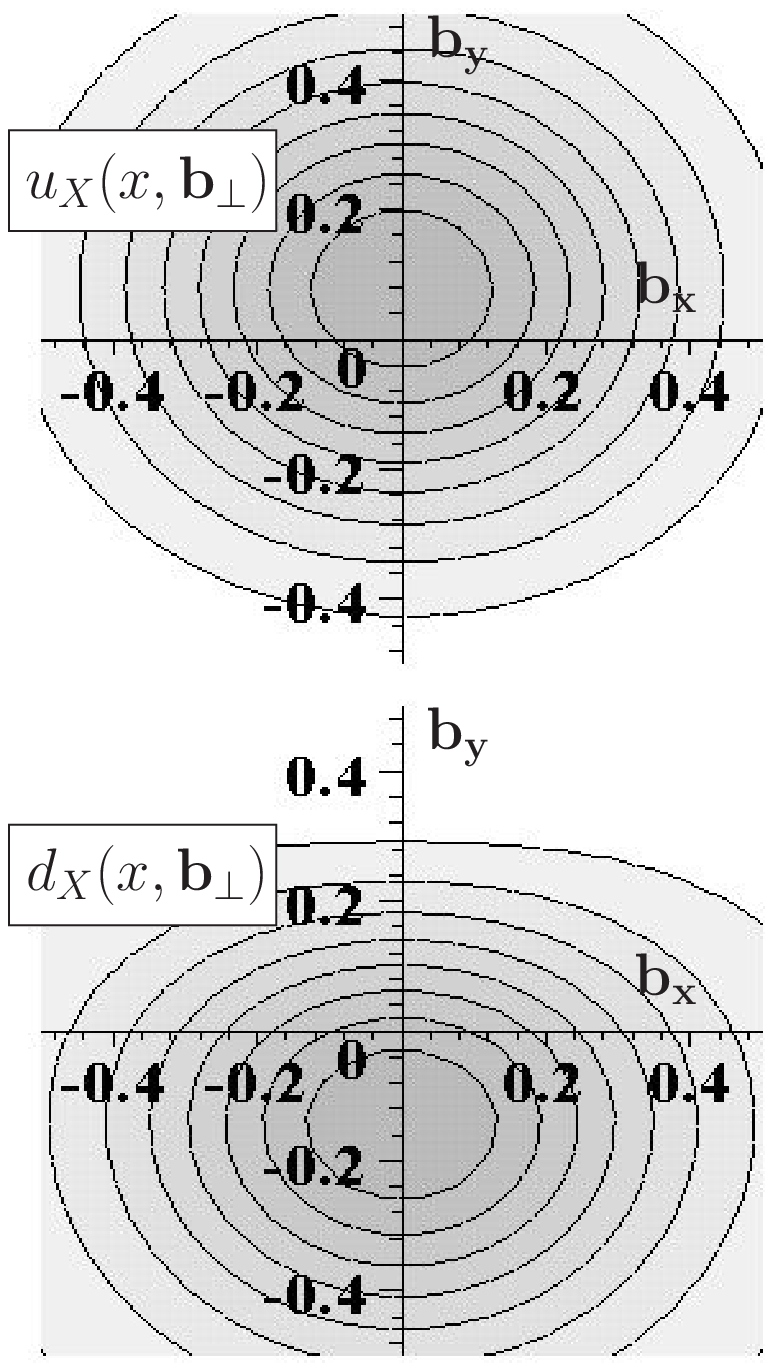}
\end{picture}
\caption{Distribution of the $j^+$ density for
$u$ and $d$ quarks in the
$\perp$ plane ($x_{Bj}=0.3$ is fixed) for a nucleon that is polarized
in the $x$ direction in the model from
Ref. \cite{IJMPA}.
For other values of $x$ the distortion looks similar.
}
\label{fig:distort}
\end{figure}  
The $\kappa_q={\cal O}(1-2)$ 
are the anomalous magnetic moment contribution
from each quark flavor to the anomalous magnetic moment of the
nucleon (with charge factors taken out), i.e.
$F_2(0) = \frac{2}{3} \kappa_u - \frac{1}{3}\kappa_d-\frac{1}{3}\kappa_s...$.
This yields $\left|d^q_y\right| ={\cal O}(0.2 fm)$, where
$u$ and $d$ quarks have opposite signs. This is a sizeable effect
as is illustrated in Fig. (\ref{fig:distort}).

The physical origin of this distaortion is that due
to the kinematics of DIS it is the $j^+=j^0+j^z$ density of the
quarks which couples to the electron: the electron in
DIS couples more strongly to quarks which move towards the
electron rather than away from it because if the quarks move towards
(collision course)
the electron the electric and magnetic forces add up, while if they
move away the electric and magnetic forces act in opposite 
directions. For relativistic particles electric and magnetic
forces are of the same magnitude. As a consequence, if the $\hat{z}$
axis is in the direction of the momentum of the virtual photon then
the virtual photon couples only to the $j^+$ component of the
quark current.

Even though the $j^0$ component
of the current density is the same on the $+\hat{y}$ and $-\hat{y}$
sides of the nucleon,
the $j^z$ component has opposite signs on the $+\hat{y}$ and 
$-\hat{y}$ sides 
if the quarks have orbital
angular momentum. Therefore the reason for the distortion is a
combination of the fact that the electron `sees' oncoming quarks
better and the presence of orbital angular momentum.

While Eq. (\ref{dipole}) is a rigorous result regarding the average
distortion of quarks with flavor $q$ relative to the center
of momentum, it still does not tell us exactly what the
density density correlations are. However, qualitatively 
we expect that the sign and magnitude
of the distortion is correlated with the sign of the 
density-density correlation.
\begin{figure}
\unitlength1.cm
\begin{picture}(10,2)(5.3,19.5)
\includegraphics{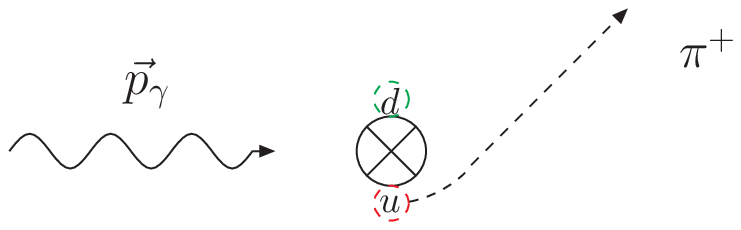}
\end{picture}
\caption{The transverse distortion of the parton cloud for a proton
that is polarized into the plane, in combination with attractive
FSI, gives rise to a Sivers effect for $u$ ($d$) quarks with a
$\perp$ momentum that is on the average up (down).}
\label{fig:deflect}
\end{figure}
Using Eq. (\ref{corel2}) we therefore expect for the resulting
Sivers effect 
\be
\langle k^y_u \rangle &<& 0 \\
\langle k^y_d \rangle &>& 0 \nonumber 
\ee
for a proton polarized in $+\hat{x}$ direction and we expect them
to be roughly of the same magnitude. 

The interpretation of these results is as follows:
the FSI is attractive and thus it ``translates'' 
position space distortions 
(before the quark is knocked out) in the $+\hat{y}$-direction into 
momentum asymmetries that favor the  $-\hat{y}$ direction and vice versa 
(Fig. \ref{fig:deflect}) \cite{mb:npa}. At least in a semi-classical 
description, this appears to be a very general observation,
which is why we expect that the signs obtained above are not
affected by higher order effects.

\section{SUMMARY}

Wilson line gauge links in gauge invariant Sivers distribution
are a formal tool to include the final state interaction in
semi-inclusive DIS experiments. The average transverse momentum
due to these Wilson lines is obtained as the correlation
between the quark density and the impulse from the spectators
on the active quark is it escapes
along its (almost) light-like trajectory.

In light-cone gauge $A^+=0$ only the gauge link at infinity 
contributes and careful regularization of the zero-modes
is necessary. However, we succeeded in expressing the net
asymmetry in terms of color density-density correlation in the
$\perp$ plane.

For a transversely polarized target the quark distribution in
impact parameter space is transversely distorted due to
the presence of quark orbital angular momentum: the $j^+$
current density is enhanced on the side where the quark orbital
motion is head-on with the virtual photon.
As the struck quark tries to escape the target, one expects on
average an attractive force from the spectators on the 
active quark, i.e. the FSI convert a left-right asymmetry for
the quark distribution in impact parameter space into a
right-left asymmetry for the $\perp$ momentum of the active quark
(Sivers effect).

The sign of the distortion in impact parameter space, and hence
the sign of the Sivers effect, for each quark flavor
is determined by the sign of the
anomalous magnetic moment contribution 
(of course with the electric charge factored out)
from that quark flavor
to the anomalous magnetic moment of the nucleon.

\end{document}